\documentclass
[a4paper,aps,prd,twocolumn,preprintnumbers,showpacs,superscriptaddress,nofootinbib]{revtex4}%
\usepackage{graphics,graphicx,epsfig}
\usepackage{amsfonts,amsmath,amssymb}
\usepackage{amsmath}
\usepackage{amsfonts}
\usepackage{amssymb}
\usepackage{graphicx}%
\begin{document}
\title{Hairy black holes sourced by a conformally coupled scalar field in $D$ dimensions}
\author{Gaston Giribet}
\email{gaston-at-df.uba.ar}
\affiliation{Departamento de F\'{\i}sica, Universidad de Buenos Aires FCEN-UBA,
IFIBA-CONICET, Ciudad Universitaria, Pabell\'on I, 1428, Buenos Aires, Argentina.}
\affiliation{Instituto de F\'{\i}sica, Pontificia Universidad Cat\'olica de Valpara\'{\i}so, Casilla 4950, Valpara\'{\i}so, Chile.}
\author{Mat\'{\i}as Leoni}
\email{leoni-at-df.uba.ar}
\affiliation{Departamento de F\'{\i}sica, Universidad de Buenos Aires FCEN-UBA,
IFIBA-CONICET, Ciudad Universitaria, Pabell\'on I, 1428, Buenos Aires, Argentina.}
\affiliation{Instituto de F\'{\i}sica de La Plata, Universidad Nacional de La Plata
IFLP-UNLP, C.C. 67, 1900, La Plata, Argentina.}
\author{Julio Oliva}
\email{julio.oliva-at-uach.cl}
\affiliation{Instituto de Ciencias F\'{\i}sicas y Matem\'{a}ticas, Universidad Austral de
Chile, Valdivia, Chile}
\affiliation{Departamento de F\'{\i}sica, Universidad de Buenos Aires FCEN-UBA,
IFIBA-CONICET, Ciudad Universitaria, Pabell\'on I, 1428, Buenos Aires, Argentina.}
\author{Sourya Ray}
\email{ray-at-uach.cl}
\affiliation{Instituto de Ciencias F\'{\i}sicas y Matem\'{a}ticas, Universidad Austral de
Chile, Valdivia, Chile}


\begin{abstract}
There exist well-known no-hair theorems forbidding the existence of hairy
black hole solutions in general relativity coupled to a scalar conformal field
theory in asymptotically flat space. Even in the presence of cosmological
constant, where no-hair theorems can usually be circumvented and black holes
with conformal scalar hair were shown to exist in $D\leq4$ dimensions, no-go
results were reported for $D>4$. In this paper we prove that these
obstructions can be evaded and we answer in the affirmative a question that
remained open: Whether hairy black holes do exist in general relativity
sourced by a conformally coupled scalar field in arbitrary dimensions. We find
the analytic black hole solution in arbitrary dimension $D>4$, which exhibits
a backreacting scalar hair that is regular everywhere outside and on the
horizon. The metric asymptotes to (Anti-)de Sitter spacetime at large distance
and admits spherical horizon as well as horizon of a different topology. We
also find analytic solutions when higher-curvature corrections ${\mathcal{O}%
}(R^{n})$ of arbitrary order $n$ are included in the gravity action.

\end{abstract}
\maketitle

\section{Introduction}

There exist well-known no-hair theorems forbidding the existence of hairy
black holes in general relativity (GR) conformally coupled to a scalar field
theory, in asymptotically flat spacetime. More precisely, in the particular
instance of a scalar field conformally coupled to GR with vanishing
cosmological constant in $D=4$ spacetime dimensions black hole solutions do
exist \cite{BBM, Bekenstein}, but they exhibit a scalar field configuration
that diverges at the horizon. In dimension $D>4$, black hole solutions of this
type simply do not exist at all \cite{griegos, Klimcik}. In the case of
non-vanishing cosmological constant, in which the no-hair theorems can usually
be circumvented, black hole solutions with conformal scalar field
configurations that are regular everywhere outside and on the horizon were
found both in $D=3$ and $D=4$ dimensions \cite{MTZ1,MTZ2, MTZ3,Anabalon};
however, the situation is quite different for $D>4$ where no-go results have
been reported \cite{Martinez}. In this paper, we prove that these restrictions
are circumvented if the scalar field theory is coupled to higher-order Euler
densities in a conformally invariant manner. In this way, we answer in the
affirmative a question that remained open: Whether hairy black holes do exist
in GR\ sourced by a conformally coupled scalar field in arbitrary dimensions.

The paper is organized as follows:\ In section II\ we introduce the
generalized conformal couplings. The theory obtained by adding such terms,
which was introduced by two of us in Ref. \cite{weones}, is the most general
theory of its sort yielding second order field equations, and hence is the
natural extension of conformal couplings to higher-dimensions. In section
III\ we present a hairy black hole solution in GR in arbitrary dimensions
$D>4$. The solution asymptote to locally (Anti-)de Sitter spacetime and admit
spherical horizon as well as horizon of a different topology. In section
IV\ we extend our solutions to the theory whose action also includes
higher-order curvature terms $\mathcal{O}(R^{n})$ for arbitrary $n$. For the
latter theory, we also find analytic black hole solutions with a backreacting
conformal scalar hair that is regular everywhere outside and on the horizon.
We briefly comment on a duality symmetry that the theory exhibits and how it
acts on the black hole solution.

\section{Conformal field theory}

We will be concerned with the special class of theory defined in Ref.
\cite{weones}, which consists of scalar matter conformally coupled to gravity
through a non-minimal coupling between a real scalar field and the
dimensionally extended Euler densities. This yields a theory whose equations
of motion are of second order and presents quite interesting properties such
as self-duality. In four dimensions this theory reduces to Einstein theory
with a conformally coupled scalar field with potential $V(\varphi
)=(\lambda/4!)\varphi^{4}$ and non-minimal coupling with the space-time
curvature $-(1/12)R\varphi^{2}$. In higher dimensions, however, the theory
admits more general couplings which can be conveniently written by defining a
four-rank tensor $S_{\ \nu\lambda\delta}^{\mu}$ made out of the Riemann
curvature tensor $R_{\ \nu\lambda\delta}^{\mu}$ and derivatives of the scalar
field. More precisely, one defines%
\begin{align}
S_{\mu\nu}^{\ \ \gamma\delta}  &  =\phi^{2}R_{\mu\nu}^{\ \ \gamma\delta}%
-4\phi\delta_{\lbrack\mu}^{[\gamma}\nabla_{\nu]}\nabla^{\delta]}\phi
+8\delta_{\lbrack\mu}^{[\gamma}\nabla_{\nu]}\phi\nabla^{\delta]}%
\phi-\nonumber\\
&  2\delta_{\lbrack\mu}^{[\gamma}\delta_{\nu]}^{\delta]}\nabla_{\rho}%
\phi\nabla^{\rho}\phi\label{uno}%
\end{align}
which, indeed, can be seen to transform covariantly under local Weyl rescaling
$g_{\mu\nu}\rightarrow\Omega^{2}g_{\mu\nu}$ and $\phi\rightarrow\Omega
^{-1}\phi$. Note that, by construction, the tensor $S_{\ \nu\lambda\delta
}^{\mu}$ has the same algebraic symmetries as those of the Riemann tensor
$R_{\ \nu\lambda\delta}^{\mu}$.

With tensor (\ref{uno}) one can easily write down the general action of the
theory as
\begin{equation}
I=\int d^{D}x\sqrt{-g}\sum_{k=0}^{\left[  \frac{D-1}{2}\right]  }\frac
{1}{2^{k}}\delta^{\left(  k\right)  }\left(  a_{k}R^{\left(  k\right)  }%
+b_{k}\phi^{D-4k}S^{\left(  k\right)  }\right)  \label{action}%
\end{equation}
where $\delta^{(0)}R^{(0)}=\delta^{(0)}S^{(0)}=1$, the symbol $\delta^{\left(
k\right)  }$ stands for the skew-symmetric Kronecker tensor%
\[
\delta^{\left(  k\right)  }=k!\ \delta_{\lbrack\alpha_{1}}^{\mu_{1}}%
\delta_{\beta_{1}}^{\nu_{1}}...\delta_{\alpha_{k}}^{\mu_{k}}\delta_{\beta
_{k}]}^{\nu_{k}},
\]
while for $k\geq0$, $R^{(k)\text{ }}$and $S^{(k)}$ represent the $4k$-rank
tensors%
\begin{equation}
R^{(k)}=\prod\limits_{r=1}^{k}R_{\quad\mu_{r}\nu_{r}}^{\alpha_{r}\beta_{r}%
}\ ,\quad S^{(k)}=\prod\limits_{r=1}^{k}S_{\quad\mu_{r}\nu_{r}}^{\alpha
_{r}\beta_{r}}\ . \label{L}%
\end{equation}

$a_{k}$ and $b_{k}$ in (\ref{action}) represent coupling constants that are in
principle arbitrary. The theory will, however, exhibit special features for
special relations between these constants.

The upper limit of the sum in (\ref{action}) is the integer part of $(D-1)/2$.
When all the couplings $b_{k}$ in (\ref{action}) vanish, the theory reduces to
Lovelock theory of gravity and, in particular, if $a_{k\neq1}=0$, it reduces
to general relativity.

The field equations coming from (\ref{action}) can be written as%
\begin{equation}
G_{\mu\nu}=T_{\mu\nu}\ ,
\end{equation}
where the symmetric tensors $G_{\mu\nu}$ and $T_{\mu\nu}$ are%
\begin{align}
G_{\mu}^{\nu}  &  =-\sum_{k=0}^{\left[  \frac{D-1}{2}\right]  }\frac{a_{k}%
}{2^{k+1}}\delta_{\mu\rho_{1}...\rho_{2k}}^{\nu\lambda_{1}...\lambda_{2k}%
}R_{\ \ \ \ \lambda_{1}\lambda_{2}}^{\rho_{1}\rho_{2}}...R_{\ \ \ \ \lambda
_{2k-1}\lambda_{2k}}^{\rho_{2k-1}\rho_{2k}}\ ,\\
T_{\mu}^{\nu}  &  =\sum_{k=0}^{\left[  \frac{D-1}{2}\right]  }\frac{b_{k}%
}{2^{k+1}}\phi^{D-4k}\delta_{\mu\rho_{1}...\rho_{2k}}^{\nu\lambda
_{1}...\lambda_{2k}}S_{\ \ \ \ \lambda_{1}\lambda_{2}}^{\rho_{1}\rho_{2}%
}...S_{\ \ \ \ \lambda_{2k-1}\lambda_{2k}}^{\rho_{2k-1}\rho_{2k}}\ .\nonumber
\end{align}

In addition, the equation for the scalar field gives%
\begin{equation}
\sum_{k=0}^{\left[  \frac{D-1}{2}\right]  }\frac{\left(  D-2k\right)  b_{k}%
}{2^{k}}\phi^{D-4k-1}\delta^{\left(  k\right)  }S^{\left(  k\right)  }=0.
\label{eqfieldarb}%
\end{equation}

The theory defined by action (\ref{action}) is the most general theory of
gravity conformally coupled to a scalar field theory yielding second order
field equations. As it should be for a conformal field with non-zero conformal
weight, the trace of its associated energy-momentum tensor, $T_{\mu}^{\nu}$,
vanishes on-shell (i.e. after Eq. (\ref{eqfieldarb}) is imposed). To
illustrate the structure of the action (\ref{action}), let us write down the
first terms explicitly: Up to terms that are linear in the Riemann tensor, the
action reads%
\begin{align}
I  &  =\int d^{D}x\sqrt{-g}(a_{0}+a_{1}R+b_{0}\phi^{D}+b_{1}\phi
^{D-2}\nonumber\\
&  \left.  (R+(D-1)(D-2)\phi^{-2}\left(  \partial\phi\right)  \right)
^{2})+\ ... \label{ssta}%
\end{align}
Then, redefining the scalar field as $\phi\rightarrow\varphi=\phi^{(D-2)/2}$,
action (\ref{ssta}) reduces to the familiar canonically normalized scalar
conformal field theory%
\begin{align}
I  &  =\int d^{D}x\sqrt{-g}\left(  \frac{1}{16\pi G}R-\frac{\Lambda}{8\pi
G}-\frac{1}{2}\left(  \partial\varphi\right)  ^{2}-\right. \nonumber\\
&  \left.  \frac{D-2}{8(D-1)}\varphi^{2}R-\frac{\lambda}{D!}\varphi^{\frac
{2D}{(D-2)}})\right)  +\ ... \label{sssta}%
\end{align}
where we have conveniently denoted $a_{0}\equiv-\Lambda/(8\pi G)$,
$a_{1}\equiv1/(16\pi G)$, and $b_{0}\equiv-\lambda/D!$. The important feature
here is that the conformal coupling $(\xi/2)\varphi^{2}R$ with $\xi
=-(D-2)/(4D-4)$ appears automatically once the kinetic term for the field
$\varphi$ is chosen to be canonically normalized, which in turn amounts to
choosing $b_{1}=-(D-2)/(8(D-1))$. This is because both terms actually come
from the same contribution $\phi^{D-4}S^{(1)}$ in (\ref{action}).

The quadratic terms in the action take the form $(a_{2}+b_{2}\phi
^{D-4})\left(  R^{2}-4R_{\mu\nu}R^{\mu\nu}+R_{\mu\nu\lambda\delta}R^{\mu
\nu\lambda\delta}\right)  +$ $...$ where the ellipses stand for terms that
contain up to second powers of the second derivative of $\phi$. This coupling
between a real scalar field and the quadratic Gauss-Bonnet term resembles the
next-to-leading contribution to the low energy effective action of string
theory. The explicit form of the $S^{(2)}$ terms can be found in
\cite{weones}. The complexity of the action$\ $increases notably\footnote{It
can, however, be somewhat simplified by resorting to the first order
formalism.} with $k$.

For the theory (\ref{action}) we will show that, remarkably, analytic hairy
black hole solutions can be found in arbitrary dimension $D>4$. As a
particular case we will find hairy black holes in GR\ coupled to scalar
conformal field theory (CFT) in asymptotically (Anti-)de Sitter spacetime
((A)dS). The existence of such solutions is remarkable because, as we
discussed above, there are stringent no-go results reported in the search for
conformal hairs in $D>4$. Our result shows which is the appropriate coupling
needed for the black hole configurations to be supported by scalar conformal matter.

\section{Hairy black holes}

Let us first consider GR coupled to the CFT defined by the matter content of
action (\ref{action}). That is, consider first the particular case $a_{k}=0$
for $k>1$ in (\ref{action}). Up to quadratic terms in the matter part, the
action reads%
\begin{align}
I  &  =\int d^{D}x\sqrt{-g}\left(  \frac{1}{16\pi G}R-\frac{\Lambda}{8\pi
G}+b_{0}\phi^{D}+b_{1}\phi^{D-4}S+\right. \nonumber\\
&  \left.  b_{2}\phi^{D-8}\left(  S_{\mu\nu\alpha\beta}S^{\mu\nu\alpha\beta
}-4S_{\mu\nu}S^{\mu\nu}+S^{2}\right)  \right)  +\ ...
\end{align}
where $S_{\mu\nu}=S_{\ \mu\rho\nu}^{\rho}$ and $S=S_{\mu}^{\ \mu}$.

Generally, in $D$ dimensions one can include $S^{(k)}$ terms for
$k=0,1,2,...[(D-1)/2]$ and there exist static black hole solutions, provided
the couplings of such terms satisfy certain relations. Their metrics take the
form%
\begin{equation}
ds^{2}=-f\ dt^{2}+f^{-1}dr^{2}+r^{2}d\Sigma_{D-2,\gamma}^{2} \label{dOne}%
\end{equation}
where $d\Sigma_{D-2,\gamma}^{2}$ is the metric of a ($D-2$)-dimensional
Euclidean space of constant curvature $\gamma=(0,\pm1)$ with \textit{radius}
one and volume $Vol_{\Sigma}$, and where the metric function $f$ is given by%
\begin{align}
f(r)  &  =\gamma-\frac{16\pi GM}{(D-2)Vol_{\Sigma}r^{D-3}}-\frac{16\pi
GQ}{(D-1)(D-2)r^{D-2}}-\nonumber\\
&  \frac{2\Lambda}{(D-1)(D-2)}r^{2}, \label{dTwo}%
\end{align}
where $a_{0}=-\Lambda/(8\pi G)$, $a_{1}=1/(16\pi G)$, $M$ is an arbitrary
constant ultimately associated to the mass of the solution, and $Q$ is given
in terms of the couplings $b_{k}$ by the relation
\begin{equation}
Q=\sum\limits_{k=0}^{\left[  \frac{D-1}{2}\right]  }\left(  D-2k-1\right)
\widetilde{b}_{k}\gamma^{k}N^{D-2k}\ , \label{dThree}%
\end{equation}
where $\widetilde{b}_{k}=b_{k}(D-1)!/(D-2k-1)!$, and where $N$ is a
dimensionful constant that appears in the scalar field configuration%
\begin{equation}
\phi\left(  r\right)  =\frac{N}{r} \label{fi}%
\end{equation}
and satisfies the following constraints
\begin{align}
\sum\limits_{k=1}^{\left[  \frac{D-1}{2}\right]  }k\ \widetilde{b}_{k}%
\gamma^{k-1}N^{2-2k}  &  =0\text{,}\label{generalrelations}\\
\sum\limits_{k=0}^{\left[  \frac{D-1}{2}\right]  }\left(  D\left(  D-1\right)
+4k^{2}\right)  \widetilde{b}_{k}\gamma^{k}N^{-2k}  &  =0\ .
\label{generalrelations2}%
\end{align}

The couplings $b_{k}$ have to obey the above constraints for the solution
(\ref{dOne})-(\ref{dTwo}) to exist in $D>4$. This implies in particular, that
in $D=5$ dimensions, for example, one needs to have $b_{0}\neq0$, $b_{1}\neq
0$, and also $b_{2}\neq0$.

Provided $\gamma\neq0$ and the metric function $f(r)$ has least one positive
root\footnote{It is easy to verify that $f$ admits positive roots for special
range of the parameter $M$ and coupling constants, so horizon does exist. For
instance, consider the example in $D=5$ ($k\leq2$), with $a_{0}>0,$ $a_{1}>0$
($\Lambda<0$), $\gamma=1,$ $b_{2}>0$ and $M$ large enough.}, solution
(\ref{dOne})-(\ref{dTwo}) represents a hairy black hole with only one
parameter, $M$. Indeed, the metric exhibits a scalar hair: The scalar field
configuration (\ref{fi}) turns out to be regular everywhere outside and on the
horizon; it only diverges at the origin, $r=0$, where the geometry develops a
curvature singularity (cf. \cite{Bekenstein, griegos, Klimcik}). The intensity
of the scalar hair, given by $N$, cannot be changed and is governed by the
quotients of coupling constants $b_{k}$. If the black hole has locally flat
horizon ($\gamma=0$), the hair disappears from the metric. In the case
$\gamma=\pm1$, in contrast, the scalar hair induces a contribution to the
gravitational potential that damps off like $\sim Q/r^{(D-2)}$, which is
consistent with the fact that matter is conformally coupled. In fact, the
fall-off condition $\sim1/r^{D}$ is the expected one for the $T_{0}^{\ 0}(r)$
component of the CFT\ stress-tensor, which is the one that couples to gravity
in the static spherically symmetric ansatz. Since the scalar field
configuration is an actual hair in the sense that it falls off more rapidly
than the Newtonian term $\sim M/r^{D-3}$, the solution happens to have a
finite mass, which is given by $M.$ The fall-off of $\phi\sim1/r$ in
(\ref{fi}) in terms of the canonically normalized field translates into
$\varphi\sim1/r^{(D-2)/2}$. All these imply that the asymptotics remains
locally (A)dS as in the non-hairy solution $\phi=0,$ $Q=0$.

Solution (\ref{dOne})-(\ref{dTwo}) shares some properties with the $D=4$ black
hole solution\footnote{Nevertheless, it is worth pointing out that the
solutions we found for $D>4$ do not exist in $D=4$. In this sense, our
solutions are not a generalization of the $D=4$ solutions, but belong to a
different class. It may be the case that there exists a larger set of
solutions to theory (\ref{action}) that would include both as special cases.}
found in Ref. \cite{MTZ3}. In the four-dimensional case, however, if
$\Lambda<0$ then the only possible horizon geometry has negative constant
curvature ($\gamma=-1$), corresponding to a topological black hole with a
Reissner-Nordstr\"{o}m type contribution to the potential. In $D>4$
dimensions, in contrast, spherical horizons ($\gamma=1$) are admitted and
scalar CFT matter contributes differently relative to gauge fields.

There is a lesson to learn from our finding, which is the following:\ When
trying to naively extend the $D\leq4$ hairy black holes to $D>4$ dimensions,
the solutions obtained do not exhibit a horizon but develop a naked
singularity \cite{Martinez}. It turns out that if what one really wants is the
horizon to persist together with a conformal scalar hair in $D>4 $, then one
has to add conformal couplings of the form $\varphi^{(2D-4k)/(D-2)}%
R^{(k)}+...$ with $k>1$ and not only the usual term $\varphi^{2}R^{(1)}$;\ the
latter is only sufficient in $D=3$ and $D=4$ where terms $S^{(k)}$ with $k>1$
do not contribute. The non-canonical kinetic terms for $\varphi$ coming from
$S^{(k)}$ are also necessary, and this is the reason why previous attempts to
find such a solution failed. Once one realizes this, the black hole solution
found above can be easily generalized even to the case where ${\mathcal{O}%
}(R^{k})$ terms for arbitrary $k$ are also included in the gravity action,
i.e. to the case $a_{k\geq2}\neq0$. Let us discuss the general case in the
next section.

\section{Higher-curvature corrections}

In $D$ dimensions, consider the action that includes both higher-curvature
terms of the form $R^{(k)}$ in the gravity Lagrangian and terms of the form
$S^{(k)}$ in the matter Lagrangian with $k>1$; namely consider the full action
(\ref{action}). It turns out that this generalized theory also admits the
following solution%
\begin{equation}
ds^{2}=-f\ dt^{2}+f^{-1}\ dr^{2}+r^{2}d\Sigma_{D-2,\gamma}^{2}\ ,
\end{equation}
where, again, the scalar field configuration is given by (\ref{fi}), while the
metric function $f\ $now fulfills the polynomial equation%
\begin{equation}
\sum\limits_{k=0}^{\left[  \frac{D-1}{2}\right]  }a_{k}\frac{(D-1)!}%
{(D-2k-1)!}\left(  \frac{\gamma-f\left(  r\right)  }{r^{2}}\right)  ^{k}%
=\frac{M(D-1)}{Vol_{\Sigma}\ r^{D-1}}+\frac{Q}{r^{D}}\ , \label{polyarb}%
\end{equation}
where $M$ is an arbitrary integration constant, and $Q$ is fixed to be%
\begin{equation}
Q=\sum\limits_{k=0}^{\left[  \frac{D-1}{2}\right]  }\left(  D-2k-1\right)
\widetilde{b}_{k}\gamma^{k}N^{D-2k}\ . \label{fixQ}%
\end{equation}

As it happens in GR, constraints (\ref{generalrelations}%
)-(\ref{generalrelations2}) between couplings have to be imposed. Equations
(\ref{generalrelations})-(\ref{generalrelations2}) fix the value of $N$ as
well the relation between the couplings $b_{k}$ of the matter part.

Equation (\ref{polyarb}) is a polinomial of degree $k$ in $f$. It gives the
solution for $f$, which can be explicitly expressed in terms of elementary
operations only for $D<10$. Expression (\ref{polyarb}) is the analogue of
Wheeler polynomial of Lovelock theory of pure gravity \cite{Wheeler}. As in
the GR\ case discussed in Section III, the solution has a single integration
constant $M$, and for generic values of the couplings in the gravity action
(i.e. $a_{k}$ being generic enough) there is always one Schwarzschild branch,
for which $M$ turns out to be the mass of the solution. Constant $Q$ is fixed
in terms of $N$ through (\ref{fixQ}), and the latter is fixed by
(\ref{generalrelations})-(\ref{generalrelations2}). Therefore, the scalar
field can be thought of as a \textit{secondary hair}, as in the case in the
three- and four-dimensional solutions previously reported in the literature.

\section{Discussion}

In this paper we have proven that hairy black hole solutions in general
relativity sourced by a conformally coupled scalar field do exist in arbitrary
dimensions. This result provides an answer in the affirmative to a question
that remained open and for which no-go results had been reported. The
backreacting solutions we found here represent geometries that asymptote to
(Anti-)de Sitter spacetime at large distances and admit both spherical
horizons and horizons of a different topology. We have also shown how such
black hole solutions extend to theories of gravity with higher-curvature
corrections ${\mathcal{O}}(R^{n})$ of arbitrary order $n$, for which we also
found exact solutions.

Before concluding, let us comment that theory (\ref{action}) exhibits other
interesting properties such as self-duality. In fact, for special relations
between couplings $a_{k}$ and $b_{k}$, the action of the theory is symmetric
under the interchange of the pure gravity part of action (\ref{action}) (i.e.
the terms $R^{(k)}$ in $I$) with the matter part of it (i.e. the terms
$S^{(k)}$ in $I$), accompanying this commutation with the inversion
$\phi\rightarrow1/\phi$ of the scalar field. The symmetry under the
interchange $R^{(k)}\leftrightarrow\phi^{D-4k}S^{(k)}$ is clearly understood
once one remembers that tensor $S_{\mu\nu}^{\ \ \gamma\delta}$ transforms
covariantly under local Weyl transformations. In fact, this symmetry
transformation can be realized by rescaling the metric as $g_{\mu\nu
}\rightarrow\phi^{2}g_{\mu\nu}$. To gain intuition and see how this duality
symmetry acts on the space of solutions of the theory one can consider the
spherically symmetric solution (\ref{dOne})-(\ref{dTwo}), (\ref{fi}) derived
above. Since, according to (\ref{fi}), $\phi=N/r$ and the duality
transformation is given by the rescaling $g_{\mu\nu}\rightarrow\phi^{2}%
g_{\mu\nu}\sim r^{-2}g_{\mu\nu}$, one finds that the black hole metric gets
transformed into a metric that is a direct product of a two-dimensional space
with coordinates ($t,r$) and the space of constant curvature $\Sigma
_{D-2},_{\gamma}$. This duality symmetry and its action on the space of
solutions of the theory will be studied in more detail in a future work. Let
us just mention here that it corresponds to a type of ultraviolet/infrared
correspondence which maps the large distance behavior of some solutions into
the short distance limit\footnote{Up to terms $R^{(1)}$ and $S^{(1)}$ this
correspondence has been explored in \cite{HAstrongweak} for AdS wave solutions
of the system in three dimensions.}.%

\[
\]
The authors thank Christos Charmousis and Mokhtar
Hassa\"{\i}ne for many useful discussions. G. G. and J. O. thank
the support of ICTP where part of this work was done.
G. G. Also thanks Jorge Zanelli and the members of CECs,
and specially thanks the Pontificia Universidad Cat\'{o}lica de Valpara\'{\i}so, where substantial part of this work was done.
The work of G. G. and M. L. was supported by grants PIP,
PICT, UBACyT, Math-AmSud, BMWF-MINCyT, from
CONICET, ANPCyT, and UBA. J. O. is supported by
the FONDECYT Grant No. 1141073. S. R. is supported by
FONDECYT Grant No. 11110176 and by CONICYT
Grant No. 791100027.

\end{document}